**Grain boundary segregation in UFG alloys processed by severe plastic deformation**

Xavier Sauvage[1], Artur Ganeev[2], Yulia Ivanisenko[3], Nariman Enikeev[2], Maxim Murashkin[2], Ruslan Valiev[2]

*1- University of Rouen, CNRS UMR 6634, Groupe de Physique des Matériaux, Faculté des Sciences, BP12, 76801 Saint-Etienne du Rouvray, France*

*2- Institute for Physics of Advanced Materials, Ufa State Aviation Technical University, K. Marx 12, Ufa 450000, Russia*

*3- Karlsruhe Institute of Technology, Institute of Nanotechnology, Hermann-von-Helmholtz-Platz 1, 76344 Eggenstein-Leopoldshafen, Deutschland*

Corresponding author: xavier.sauvage@univ-rouen.fr

**Abstract**

Grain boundary segregations were investigated by Atom Probe Tomography in an Al-Mg alloy, a carbon steel and Armco® Fe processed by severe plastic deformation (SPD). In the non-deformed state, the GBs of the aluminium alloy are Mg depleted, but after SPD some local enrichment up to 20 at.% was detected. In the Fe-based alloys, large carbon concentrations were also exhibited along GBs after SPD. These experimental observations are attributed to the specific structure of GBs often described as "non-equilibrum" in ultra fine grained materials processed by SPD. The grain boundary segregation mechanisms are discussed and compared in the case of substitutional (Mg in fcc Al) and interstitial (C in bcc Fe) solute atoms.

# 1. Introduction

Using severe plastic deformation (SPD) techniques, it has been shown that ultrafine grained (UFG) structure could be achieved in various metals and metallic alloys [1]. The grain size refinement mechanism during SPD is controlled by the generation of dislocations and dynamic recovery processes, i.e. the way dislocations do dynamically reorganize to form first low angle boundaries (dislocation cell walls) and finally, for larger strains, high angle boundaries [1]. These features may explain why a small change in the Mg concentration in aluminum can dramatically change the grain size achievable by SPD. Indeed, if the typical grain size achieved by High Pressure Torsion (HPT) in pure Al is about 800 nm [2], it drops down to only 150 nm in Al-3%Mg [3-5]. This trend has been confirmed by a large number of experimental investigations [6-9]. Beyond the strong interactions of Mg with dislocations, it is also well known that this element is prone to grain boundary (GB) segregation. Both experimental evidence [10, 11] and simulation work [12-14] have demonstrated this tendency.

GB segregation is an extremely sensitive phenomenon because it may affect significantly the material properties like the corrosion resistance, the mechanical behavior or the thermal stability [15-17]. In UFG materials the proportion of grain boundaries is much larger than in conventional coarse-grained alloys. Beside, GB boundaries achieved by SPD usually exhibit special features, like long range elastic stresses and enhanced free volume [1, 19]. Thus they are sometimes called "non-equilibrium" GB and they do appear as ideal candidate for segregations that may minimize their local energy state [17, 18]. There is some indirect evidence of grain boundary segregation in UFG materials processed by SPD. For example it has been demonstrated that the thermal stability of Ni [20] or Ag [21] is linked to the amount of impurities (supposed to segregate at GBs). A similar way, the UFG structure of Cu processed by SPD is more stable upon annealing if the material is alloyed with a small amount of Sb [22]. However, one should note that the role of solute elements on the thermal stability of nanocrystalline materials could be rather complex and not necessarily linked to segregations [23]. But using Atom Probe Tomography (APT) some direct proofs of GB segregation were obtained in various UFG alloys processed by SPD, like steels [24, 25], Ti alloys [26], Ni [27] and Al alloys [28-32]. It is worth noticing that GB segregations might be a key issue for the application of UFG alloys, since it could significantly affect the properties [15] like the strength [32-35] or the thermal stability [20-22].

In this paper, SPD induced GB segregation in UFG materials is discussed on the basis of new experimental data collected on two different systems, namely Al-Mg and Fe-C, where solute atoms are respectively located on substitution and interstitial sites in the matrix. These data were collected by APT to reveal local information about chemical gradients. For the Al-Mg system, some attempts were made to compare the distribution of Mg atoms in the annealed material (coarse grained) including along GBs, and in the UFG state (after SPD) were GBs are different and described as "non-equilibrum" [1, 19]. For the Fe-C system, two different steels were investigated to find out if the amount of available solute element could significantly affect the segregation level at GBs.

## 2. Experimental

The Al-Mg alloy investigated in the present study is a hot-pressed commercial alloy containing 5.7% Mg, 0.32%Sc and 0.4%Mn (wt.%). Before grain refinement by SPD, the material was homogenized at 380°C during 2h (above the solubility limit of Mg in fcc Al), the grain size being larger than 10μm. Discs (25 mm in diameter and 2mm in thickness) were processed High Pressure Torsion (HPT) at room temperature under a pressure of 6 GPa up to 20 revolutions.

Two different Fe-based materials were also investigated. A commercial Armco ® iron with a very low level of impurity and a commercial C45 steel with a nominal carbon concentration of 0.45wt.%. Before grain refinement by SPD, the C45 steel was patented as following: homogenization treatment at 900°C during 1 hour and then transformed at 500°C to obtain a pseudo-pearlitic structure with nanoscaled $Fe_3C$ particles. Both Fe-based materials were processed by HPT at 350°C under a pressure of 6GPa, up to 5 revolutions (discs 10 mm in diameter and 0.3 mm in thickness).

APT samples were prepared by standard electropolishing techniques at a distance of 6mm (respectively 3mm) from the disc center for the Al-Mg alloy (respectively Fe-based materials). Analyses were performed in UHV conditions, using an energy compensated atom probe equipped with an ADLD detector [36]. Two different configurations were used: a standard reflectron (small field of view) or a curved reflectron (large field of view). Samples were field evaporated using electric pulses (30kHz pulse repetition rate and 20% pulse fraction) at a temperature of 80K. Such conditions are known to provide quantitative measurements in Fe-C alloys (the carbon quantification was made following the procedure described in [24]) but not in Al-Mg. Indeed, at a temperature higher than 40K a significant amount of Mg might be evaporated between pulses (preferential evaporation) leading to non-quantitative measurements. However, SPD samples being extremely sensitive and prone to early specimen failure during analysis, it was impossible to analyze them at a temperature below 80K. Thus, a systematic correction was applied on the Mg content from a calibration measurement performed on a unique sample of the non deformed material (7.0 ±0.2 at.%Mg measured at 40K vs 4.4 ±0.2 at.% at 80K).

## 3. Results

### 3.1 Al-Mg alloy

*Mg distribution before deformation*

Most of APT data set collected for the non-deformed material (homogenized at 380°C during 2h) do exhibit a very homogeneous distribution of Mg in solid solution as expected. The average concentration measured is about 7.0 ±0.2 at.%, very close to the nominal composition. However, as shown in Fig. 1 (a), a planar defect, depleted in Mg was also intercepted. No crystallographic information is available in this data set, but this feature is most probably a GB. A composition profile was computed across (Fig. 1 (b)), and considering the thickness of the sampling volume (1nm), it turns out that this GB is characterized by a 2nm thick layer depleted in Mg (sharp gradients and local concentration almost down to zero). Since the alloy does contain some Sc, it was not surprising to find some spherical $Al_3Sc$ nanoscaled particles (Fig. 1 (c)). Their size is in a range of 5 to 10nm and the amount of Sc remaining in solid solution in the aluminum matrix is below the detection limit of the instrument (i.e. below 50 ppm). One should note that the small black dots corresponding to Sc atoms in the reconstructed volume of Fig. 1 (c) are related to the detection noise in the mass range of Sc ions in the mass spectrum. The alloy also contains some Mn but this element was only detected in large Al-Mn intermetallic particles (data not shown here). Such a Mg depletion along GBs is not consistent with recent simulation work showing that a Mg rich layer should cover GBs in fcc Al [12-14]. It is however important to note that only very specific grain boundaries were simulated in these calculations. Beside, only one GB in the coarse grained state was successfully analyzed in the present study. Paine and co-authors [11] have performed a systematic study of GB segregation by STEM-EDX in a similar alloy (6.5

wt.%Mg) homogenized at 350°C during 30min. They measured that 8 out of 11 GBs were significantly depleted in Mg, thus it is reasonable to think that this only APT measurement is rather representative of the average GB state before SPD.

*Mg distribution in the UFG structure resulting from SPD*

As detailed in our previous work [9], after HPT at room temperature, the average grain size is about 130nm. There are still $Al_3Sc$ particles and TEM observations (not shown here) have revealed that they are often located along GBs, pinning them and probably limiting their mobility. Such particles were also detected by APT (Fig. 3 (a)), but the most interesting is the distribution of Mg atoms that now appears locally absolutely non uniform, confirming our preliminary observations [32]. In the data set displayed in Fig. 3; a GB boundary, pinned by an $Al_3Sc$ particle, was identified (arrowed). A profile computed across this boundary shows that there is a significant drop of the apparent local density in the 3D reconstructed volume across this boundary. This is a typical feature resulting from the misorientation of the two grains located on each side of the boundary. Indeed, the evaporation field may slightly change as a function of the crystallographic orientation, leading to some small change in the local magnification and thus in the local density. The local Mg enrichment is relatively weak along the boundary but strong Mg segregations clearly appear in the vicinity of this boundary. The small volume selected in this specific region (Fig. 3 (c)) clearly shows a kind of cell structure. Cells seem almost completely depleted in Mg while their boundaries appear fully covered. A 2D Mg chemical map was computed to image the concentration gradients in this area (Fig. 3 (d)). It clearly reveals that along the cell boundaries the average concentration is about 10at.%Mg, with maxima as high as 20at.%. Other volumes with a more uniform distribution of Mg were also analyzed (not shown here). The average composition is lower than in the non-deformed state (6.4±0.5 vs 7.0±0.2 at.%) and is consistent with Mg segregations along GBs or in their vicinity.

Such large local fluctuations of the Mg content could also be attributed to some Al-Mg intermetallic precipitates even if the local concentration never reaches the 40at.% Mg required for the nucleation along GB of the equilibrium $Al_3Mg_2$ phase [37]. Indeed, in the Al-Mg system, the nucleation of GP zones or metastable phases with a $L1_0$ or $L1_2$ structure and a Mg concentration of 25at.% (i.e. close to the local concentrations measured in the present investigations) have been reported in the literature [38-41]. Beside, one should note that the dynamic precipitation of GP zones during plastic deformation of an Al-Cu alloys was recently observed [41], and the nucleation of metastable precipitates in an Al-Mg-Si-Cu alloy processed by ECAP at 200°C was also reported [29]. However, it was shown that metastable intermetallic phases and GP zones in the Al-Mg system are unstable at room temperature or higher if the Mg content of the alloy is below 9at.% [38-41], like in the present material. Moreover, second phase particles and precipitates are known to dissolve during SPD at room temperature (sheared by dislocations) [42, 43]. Thus, the high concentration of Mg locally measured by APT can reasonably be attributed without ambiguity to segregations along crystalline defects.

The SPD induced GB segregation of Mg in a similar alloy was already reported in a previous work [32], but here new data showing a more complex situation are provided. A GB pinned by an $Al_3Sc$ particle was intercepted but most of the segregated Mg is located in a complex network in the vicinity of the boundary (Fig. 2). Moreover, the analysis of a Mg depleted grain boundary in the annealed material (Fig. 1) unambiguously prove that such features are typical of UFG structures resulting from the SPD process. Thus, it seems that there is no systematic Mg enrichment of GBs during SPD, but a systematic flow of Mg toward the GBs induced by the plastic deformation. Beside, the more or less complex distribution of Mg in the GB area might be related to local distortions, i.e. to the non equilibrium state of the specific boundaries created during HPT [1, 19].

**3.2 Fe-based materials**

*<u>Carbon distribution in the UFG C45 steel processed by HPT</u>*

Before HPT, as a result of the patenting treatment, carbon atoms are well partitioned in the C45 steel, most of them being in $Fe_3C$ carbides while a very small amount, typically less than 0.1%, is in solid solution in the bcc ferrite matrix. After HPT at 350°C, ultrafine grains are elongated with an aspect ratio of about 3 (mean length 250 nm and width 70 nm, estimated from TEM data that are not shown here). Beside, as shown in the 3D reconstructed volume displayed on the Fig. 3 (a), the carbon distribution has changed. Both nanoscaled and elongated carbides (about 20 nm in length) are clearly exhibited together with some diffuse segregations along some boundaries. A composition profile computed across the carbide/matrix interface (Fig. 3 (b)) clearly indicates that these carbides are partly dissolved as reported in some earlier work [24, 25]. Indeed, the carbon concentration is well below 25 at.% as expected for stoichiometric cementite ($Fe_3C$ carbide), it is only in a range of 15 to 20 at.%. Following this partial decomposition, carbon atoms released in the matrix do segregate along defects like boundaries where the local concentration is up to 2 at.% (see the concentration profile Fig. 3 (c)). These observations are fully consistent with our previous studies about the contribution of such GB segregation on the grain size refinement mechanism in a high carbon steel processed by SPD [24]. Beside, we show as well that these segregations could be directly connected to some carbides. It seems indeed that some of them pin the new boundaries created during SPD and then directly "feed" them with solute elements. The interfacial excess of carbon at the boundary was estimated from the APT data and is in a range of 3 to 5 at/nm$^2$. The segregation of carbon along defects like dislocations or GBs in steel is a well known feature. Takahashi and co-authors have very recently investigated such interfacial excess of carbon as a function of the GB misorientation angle [44], showing that larger misorientation leads to higher level of segregation. The value measured in the present study is consistent with data reported for a GB misorientation angle of about 30°.

*<u>Carbon distribution in the UFG Armco ® iron processed by HPT</u>*

The Armco® iron does contain a very low amount of impurity. APT analyses, performed on five samples, providing each of them several millions of atoms, have shown that the following elements could be detected in solid solution: Al (50 to 100 ppm), P (50 to 100 ppm), C (100 to 200 ppm) and N (500 to 800 ppm). After HPT at 350°C, the grain size is in a range of 200 to 250 nm (estimated from TEM data not shown here), and as shown on Fig. 4 (a), carbon atoms are not homogeneously distributed. Some segregations along planar defects are indeed clearly revealed in the 3D reconstructed volume. On this image, for a better visualization of the segregation along a boundary, carbon atoms were artificially represented with dots much bigger than Fe atoms. The composition profile computed across the boundary (direction indicated by the arrow), shows that the local carbon concentration is up to about 0.5 at.% in a thin layer spreading on each side of the boundary over a distance of about 2nm. Thus, some SPD induced segregations occurred during the HPT process even if the amount of C is close to the solubility limit in bcc iron at 350°C [45]. One should note that no significant concomitant segregation of N was detected, in agreement with other published data [44].

The interfacial excess of carbon at the boundary estimated from the APT is in a range of 0.3 to 0.5 at/nm$^2$. This is much lower than in the C45 steel processed by HPT in similar conditions, and this is obviously attributed to the lower amount of carbon available. This interfacial excess is also four times lower than the data reported for a steel containing only 15 at. ppm of carbon (i.e. one order of magnitude lower than the present material) [44]. However, in the present case, the grain size is much smaller (0.2 μm vs 20 μm) and it can easily be estimated that one third of carbon atoms are segregated along GBs while it was only one out of seven in the annealed and fully recrystallized steel investigated by Takahashi and co-authors. Thus, the carbon segregation along GBs is clearly promoted by the severe plastic deformation.

## 4. Discussion about the SPD induced GB segregation

The mechanisms of plastic deformation in UFG materials are not controlled by dislocation storage inside grains that are often almost completely free of internal substructures. Most of dislocations are indeed emitted from GBs, and then immediately propagate through the grain's interior up to opposite GBs where they are absorbed and dynamically recover. Thus, there is a constant flow of dislocations toward GBs. Beside, it was demonstrated also that a large density of point defects and especially vacancies are created [46-52]. Like in irradiated materials, GBs are sinks for these vacancies (some might also be annihilated along dislocations), which leads also to a constant vacancy flux toward them. Beside, it has been demonstrated that Mg atoms in solid solution in fcc Al are strongly coupled to vacancies, with a positive binding energy in a range of 0.2 to 0.4 eV [53-55]. Thus, the atomic scale mechanism of the SPD induced segregation observed in the present study, is probably very similar to that of non-equilibrium segregation resulting from irradiation [15, 56]. Mg being strongly coupled to vacancies, the flux of these point defects toward GBs drags Mg atoms during the plastic deformation, leading to strong local enrichments. At room temperature, dynamic recovery processes of dislocations are limited, leading to the formation of the so-called "non equilibrium grain boundaries" with large local distortions (due to extrinsic dislocations) [1, 19]. Mg atoms being larger than Al, some elastic strain could be relaxed by Mg clusters giving rise to local compressive stresses. Thus it is reasonable to think that Mg segregation may stabilize GBs by reducing their energy [16, 17]. In the present work, large local enrichment in Mg was also observed in a network in the vicinity of a boundary. One should note that in aged dilute Al-Mg alloys, it was demonstrated that Mg atoms easily segregate along dislocation loops [53]. Similar features have been reported in irradiated materials where the typical loop size resulting from the agglomeration of point defects is in a range of 10 to 20 nm [56]. It is interesting to note that this value is close to the length scale of the network imaged in Fig. 3. However, one cannot exclude that the Mg mobility could also be enhanced by dislocations. Pipe diffusion mechanism along dislocation cores [57] may operate but also solute drag [58]. Indeed, moving dislocations are absorbed at GB during SPD, which leads to a constant dislocation flow toward the boundary. Due to the enhanced atomic mobility of Mg promoted by SPD induced vacancies, these dislocations could drag a significant amount of Mg even at room temperature. Beside, unusually fast diffusion of Mg was observed in a UFG Al-Mg alloy [59] and more generally fast diffusion paths have recently been reported in SPD materials [60].

Carbon atoms in solid solution in the bcc Fe are located on interstitial sites, while Mg atoms in the fcc Al are on substitutional sites, leading to different diffusion mechanisms. However, carbon atoms are also known to strongly interact with dislocations leading to strong segregations [42, 61, 62], and also to be strongly bonded to vacancies forming the so called C-V complexes [63, 64]. Thus, the segregation kinetic of this interstitial element (i.e. the flow of C atoms towards the grain boundaries during SPD), is also most probably related to the annihilation of vacancies and dislocations in the vicinity of GBs.

## 5. Conclusions

i) The results of this work demonstrate that SPD induced grain refinement in metallic alloys may also leads to the active formation of GB segregations of solute elements or impurities. These segregations may exhibit specific morphologies and unusually high local concentrations that were not observed before in metallic materials subjected to lower levels of plastic deformation and/or conventional thermal treatments.

ii) The formation of such segregations in SPD materials is attributed to the generation of "non-equilibrium" grain boundaries, which contain excessive density of extrinsic dislocations and thus additional local lattice strains and enhanced free volumes.

iii) There are experimental evidences that such SPD induced GB segregations may improve the material's properties. Therefore, it is believed that there is an opportunity to control such GB local features and thus to tune some properties of UFG alloys via SPD processing parameters.

**Acknowledgements**

One of the authors, Yu Ivanisenko, is thankful to DFG for funding support of a part of this investigation. The authors (A.Ganeev, R.Valiev and M. Murashkin ) are thankful to RFBR – 11-08-91330 for partial support of the investigation.

**References**

[1]   R.Z. Valiev, R.K. Islamgaliev, I.V. Alexandrov, *Prog. Mater. Sci.* **2000,** *45*, 103.

[2]   A.P. Zhilyaev, K. Oh-ishi, T.G. Langdon, T.R. McNelley, *Mater. Sci. Eng. A* **2005**, *410-411*, 277.

[3]   Z. Horita, T.G. Langdon, *Mat. Sci. Eng.* **2005**, *410-411*, 422.

[4]   G. Sakai, Z. Horita, T.G. langdan, *Mat. Sci. Eng. A* **2005**, *393*, 344.

[5]   D.G. Morris, M.A. Munoz-Morris, *Acta Mater* **2002**, *50*, 4047.

[6]   T. Morishige, Y. Hirata, T. Uesugi, Y. Takigawa, M. Tsujikawa and K. Higashi, *Scripta Mater* **2011**, *64*, 355.

[7]   Y. Iwahashi, Z. Horita, M. Nemoto, T.G. Langdon, *Metall Mat Trans A* **1998**, *29*, 2503

[8]   P.B. Prangnell, J.R. Bowen, P.J. Apps, *Mat Sci Eng A* **2004**, *375-377*, 178.

[9]   M.Yu. Murashkin, A.R. Kilmametov, R.Z. Valiev, *Phys. Met. Metall.* **2008**, *106*, 90.

[10]  T. Malis, MC.C Chaturvedi, *Jour Mat Sci* **1982**, *17*, 1479

[11]  D.C. Paine, G.C. Weatherly, K.T. Aust, *Jour Mat Sci* **1986,** *21*, 4257

[12]  X.-Y. Liu, J.B. Adams, *Acta Mater* **1998**, *46*, 3467

[13]  X. Liu, X. Wang, J. Wang, H. Zhang, *J. Phys: Condens. Matter* **2005**, *17*, 4301.

[14]  V.I. Razumovskiy, A.V. Ruban, I.M. Razumovdkii, A.Y. Lozovoi, V.N. Butrim, Yu. Kh. Vekilov, *Scripta Mater* **2011**, *65*, 926.

[15]  P. Lejcek, *"Grain boundary segregations in metals",* Springer-Verlag Berlin Heidelberg **2010**.

[16]  R. Kirchheim, *Acta Mater* **2007**, *55*, 5129.

[17]  R. Kirchheim, *Acta Mater* **2007**, *55*, 5139.

[18]  R.Z. Valiev, *Mater. Sci. Forum* **2008**, *584-586*, 22.

[19]  X. Sauvage, G. Wilde, S.V. Divinski, Z. Horita, R.Z. Valiev, *Mat Sci Eng A* **2012**, *540*, 1

[20]  H.W. Zhang, X. Huang, R. Pippan, N. Hansen, *Acta Mater.* **2010**, *58*, 1698.

[21] Z. Hegedus, J. Gubicza, M. Kawasaki, N.Q. Chinh, Z. Fogarassy, T.G. Langdon, *Mat Sci Eng A* **2001**, *528*, 8694.

[22] R.K. Rajgarhia, A. Saxena, D.E. Spearot, K.T. Hartwig, K.L. More, E.A. Kenik, H. Meyer, *J. Mater. Sci.* **2010**, *45*, 6707.

[23] C. C. Koch, R. O. Scattergood, K. A. Darling, J. E. Semones, *J. Mater Sci* **2008**, *43*, 7264.

[24] X. Sauvage, Y. Ivanisenko, *J. Mater. Sci.* **2007**, *42*, 1615.

[25] S. Ohsaki, K. Hono, H. Hidaka, S. Takaki, *Scripta Materialia* **2005**, *52*, 271.

[26] I. Semenova, G. Salimgareeva, G. Da Costa, W. Lefebvre, R. Valiev, *Adv. Eng. Mater.* **2010**, *12*, 803.

[27] X. Sauvage, G. Wilde, R. Valiev, *Mat. Sci. Forum* **2011**, *667-669*, 169.

[28] G. Nurislamova, X. Sauvage, M. Murashkin, R. Islamgaliev, R. Valiev, *Phil. Mag. Lett.* **2008**, *8*, 459.

[29] X. Sauvage, M. Yu. Murashkin, R.Z. Valiev, *Kovove Materialy-Metallic Materials* 2**011,** *49*, 11.

[30] P. V. Liddicoat, X-Z. Liao, Y. Zhao, Y.T. Zhu, M.Y. Murashkin, E.J. Lavernia, R.Z. Valiev, S.P. Ringer, *Nature Com.* **2010**, *1*, 63.

[31] G. Sha, Y.B. Wang, X.Z. Liao, Z.C. Duan, S.P. Ringer, T.G. Langdon, *Acta Mater.* **2009**, *57*, 3123.

[32] R. Z. Valiev, N.A. Enikeev, M. Yu. Murashkin, V.U. Kazykhanov, X. Sauvage, *Scripta Mater.* **2010**, *63*, 949.

[33] N.Q. Vo, J. Schäfer, R.S. Averback, K. Albe, Y. Ashkenazy, P. Bellon, *Scripta Mater* **2011**, *65*, 660.

[34] X.F. Zhang, T. Fujita, D. Pan, J.S. Yu, T. Sakurai, M.W. Chen, *Mat. Sci. Eng. A* **2010**, *527*, 2297.

[35] J. Schäfer, A. Stukowski, K. Albe, *Acta Mater* **2011**, *59*, 2957.

[36] G.D. Costa, F. Vurpillot, A. Bostel, M. Bouet, B. Deconihout, *Rev. Sci. Instrum.* **2005**, *76*, 013304.

[37] R. Goswami, G. Spanos, P.S. Pao, R.L. Holtz, *Mat. Sci. Eng. A* **2010**, *527*, 1089.

[38] M.J. Starink, A.-M. Zahra, *Philos Mag A* **1997**, *73*, 701.

[39] K. Osamura, T. Ogura, *Met Trans A* **1984**, *15*, 835.

[40] T. Sato, Y. Kojima, T. Takahashi, *Met Trans A* **1982**, *13*, 1373.

[41] W.Z. Han, Y. Chen, A. Vinogradov, C.R. Hutchinson, *Mat Sci Eng A* **2011**, *528*,

7410.

[42] M. Murayama, Z. Horita, K. Hono, *Acta mater.* **2001**, *49*, 21

[43] X. Sauvage, X. Quelennec, J.J. Malandain, P. Pareige, *Scripta Mater* **2006**, *54*, 1099

[44] J. Takahashi, K. Kawakami, K. Ushioda, S. Takaki, N. Nakata, T. Tsuchiyama, *Scripta Mater* **2012**, *66*, 207.

[45] C.A. Wert, *Journal of Metals* **1950**, *188*, 1242.

[46] X. Quelennec, A. Menand, J.M. Le Breton, R. Pippan, X. Sauvage, *Philos. Mag.* **2010**, *90*, 1179.

[47] D. Setman, E. Schafler, E. Korznikova and M. Zehetbauer, *Mat. Sci. Eng. A* **2008**, *493*, 116.

[48] S. Van Petegem, F. Dalla Torre, D. Segers and H. Van Swygenhoven, Scripta Mater. **2003**, *48*, 17.

[49] R. Würschum, W. Greiner, R.Z. Valiev, M. Rapp, W. Sigle, O. Schneeweiss and H. Schaefer, *Scripta Metall.* **1991**, *25*, 2451.

[50] G. Saada, *Physica* **1961**, *27*, 657.

[51] A.L. Ruoff and R.W. Balluffi, *J. Appl. Phys.* **1963**, *34*, 2862.

[52] H. Mecking and Y. Estrin, *Scripta Metall.* **1980**, *14*, 815.

[53] J.D. Embury, R.B. Nicholson, *Acta Metall* **1963**, *11*, 347.

[54] C.R.S. Beatrice, W. Garlipp, M. Cilense, A.T. Adorno, *Scripta Metall Mater* **1995**, *32*, 23.

[55] C. Panseri, F. Gatto, T. Federighi, *Acta Metall* **1958**, *6*, 198.

[56] A. Etienne, B. Radiguet, N.J. Cunningham, G.R. Odette, P. Pareige, *Jour of Nuc Mat* **2010**, *406*, 244.

[57] R.C. Picu, D. Zhang, *Acta Mater* **2004**, *52*, 161.

[58] A. Van den Beukel, *Phys. Status Solidi A* **1975**, *30*, 197.

[59] T. Fujita, Z. Horita, *Philos Mag A* **2002**, *82*, 2249.

[60] G. Wilde, J. Ribbe, G. Reglitz, M. Wegner, H. Rösner, Y. Estrin, M. Zehetbauer, D. Setman, S. Divinski, *Adv. Eng. Mater.* **2010**, *12*, 758.

[61] Wilde J, Cerezo A, GDW Smith, *Scripta Mater* **2000**, *43*, 39.

[62] V.N. Gridnev, V.G. Gavrilyuk, *Phys Metals* **1982**, *4*, 531.

[63] A. Vehanen, P. Hautojarvi, J. Johansson, J. Yli-Kauppila, *Phys. Rev. B* **1982**, *25*, 762.

[64] J.E. McLennan, *Acta Metall* **1965**, *13*, 1299.

**Figure captions**

**Figure 1**
APT data collected for the Al-Mg alloy before HPT processing. (a) 3D reconstructed volume showing a thin Mg depleted layer attributed to a GB (only Mg atoms are displayed for more clarity) ; (b) Concentration profile computed across the Mg depleted layer (sampling volume thickness 1nm) ; (c) 3D reconstructed volume showing the homogeneous distribution of Mg in the grain interior and a Sc rich nanoscaled particle.

**Figure 2**
APT data collected for the Al-Mg alloy processed by HPT. (a) 3D reconstructed volume showing the heterogeneous distribution of Mg in the vicinity of a GB pinned by an $Al_3Sc$ particle; (b) Mg concentration profile and evolution of the apparent density across the GB (thickness of the sampling volume 1nm); (c) small selection showing in a more appropriate orientation the heterogeneous distribution of Mg in the vicinity of the GB; (d) 2D Mg concentration map corresponding to the small selection and showing that the local Mg concentration is up to 20at.%

**Figure 3**
APT data collected on the C45 steel processed by HPT. (a) 3D reconstructed volume showing the heterogeneous distribution of carbon atoms (red dots), both carbides and segregations along boundaries are exhibited. (b) Carbon concentration profile across a carbide/matrix interface showing that $Fe_3C$ carbides are partly dissolved leading to a significant carbon release in the ferritic matrix (direction labeled A on the image (a) sampling volume thickness 2nm). (c) Carbon concentration profile across a segregation along a boundary showing that the local concentration is up to about 2at.% (direction labeled B on the image (a) sampling volume thickness 2nm).

**Figure 4**
APT data collected on the Armco iron processed by HPT. (a) 3D reconstructed volume showing the heterogeneous distribution of carbon atoms (red dots), some segregations along boundaries are clearly exhibited. (b) Carbon concentration profile across a segregation along a boundary (see arrow on image (a)) showing that the local concentration is up to about 0.5at.% (sampling volume thickness 1nm)

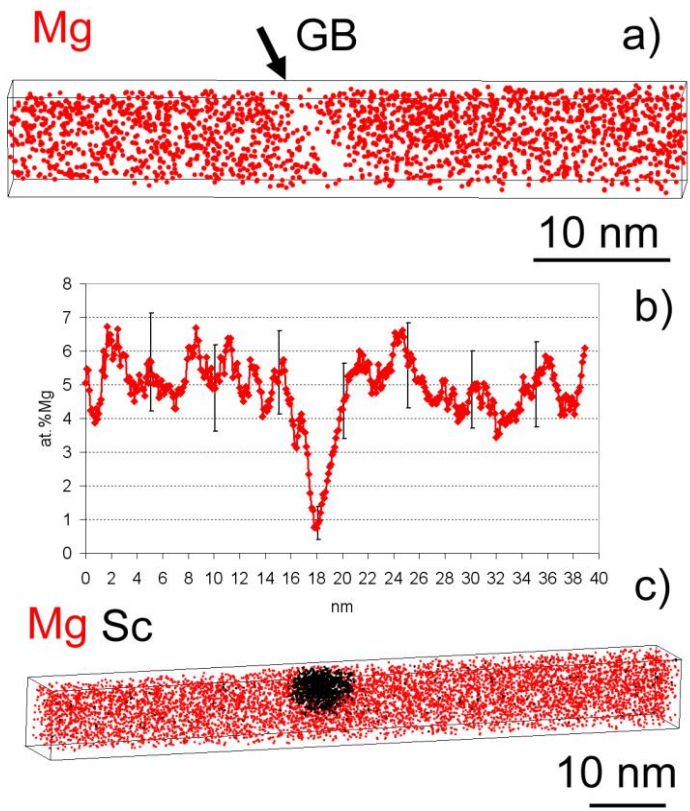

Figure 1

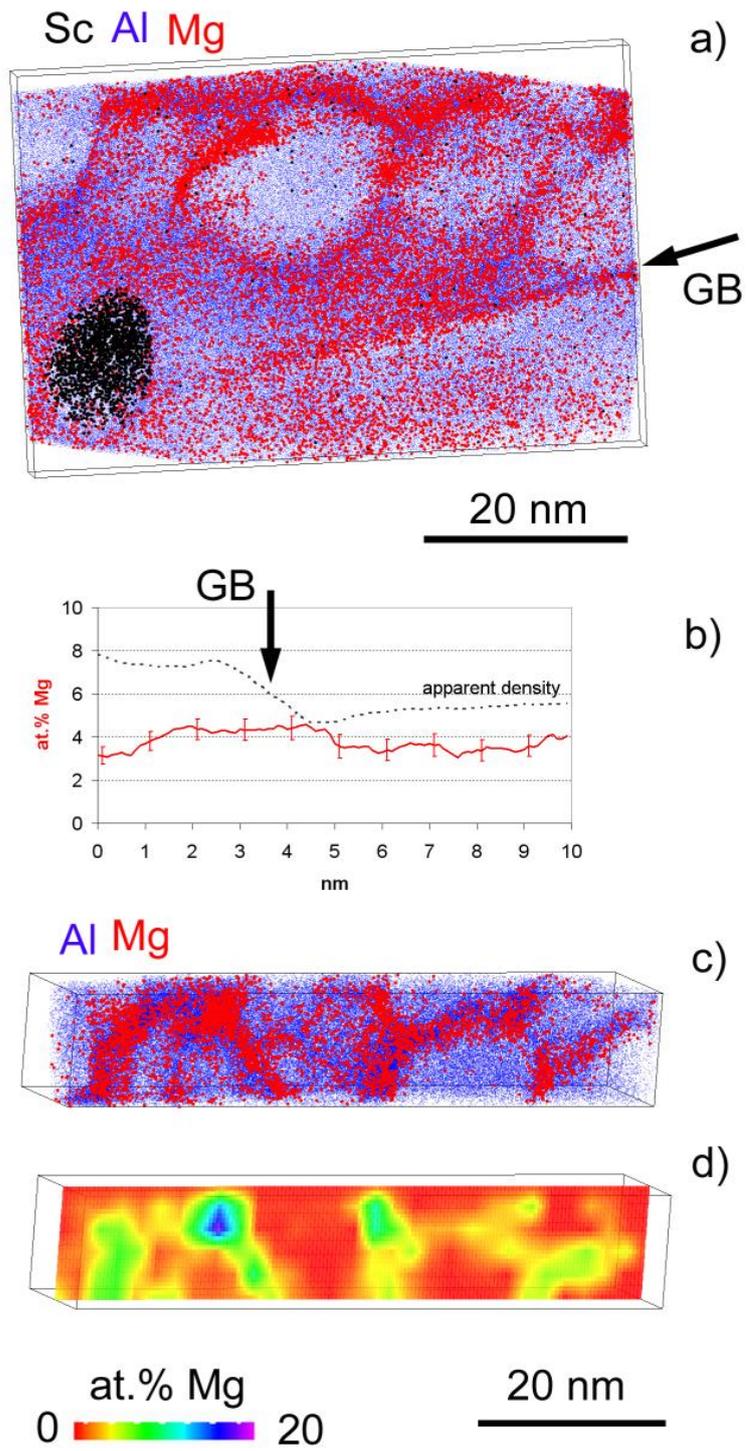

Figure 2

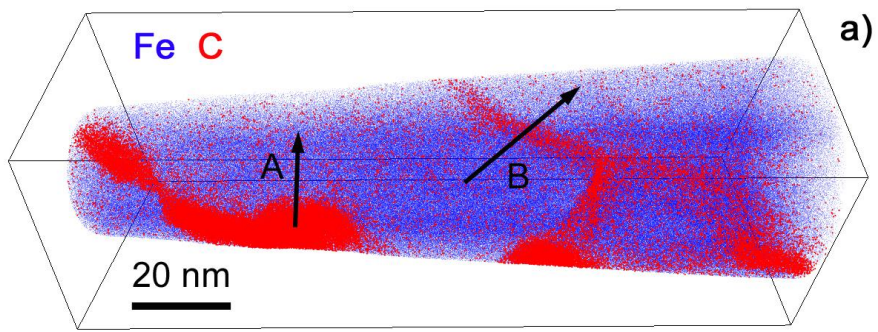

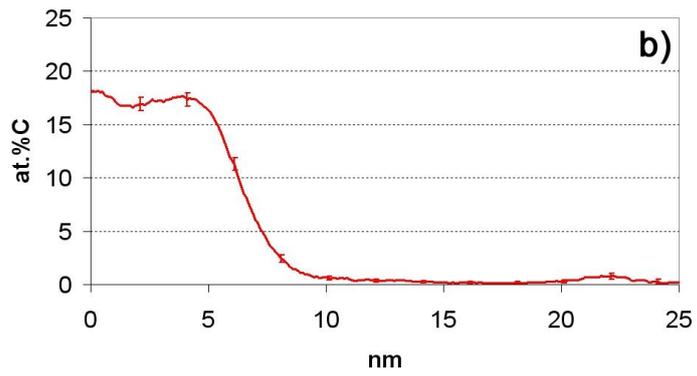

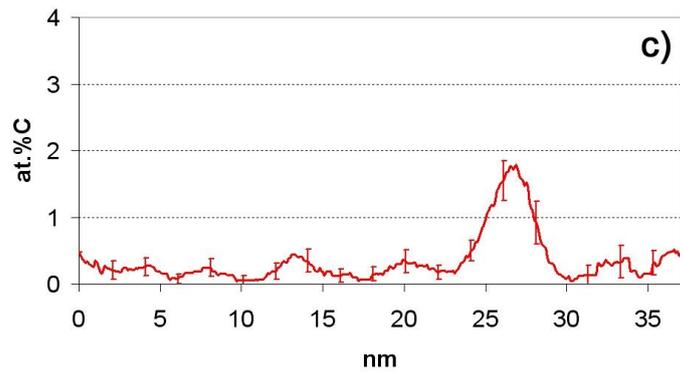

Figure 3

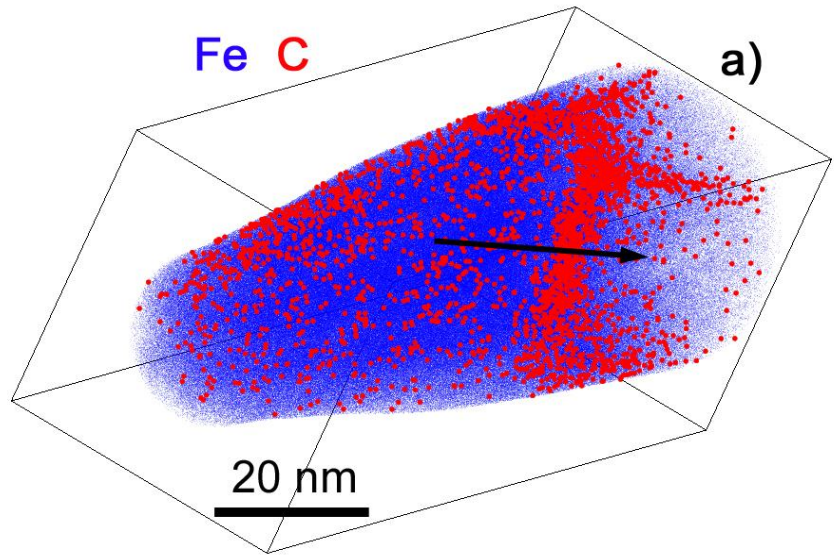

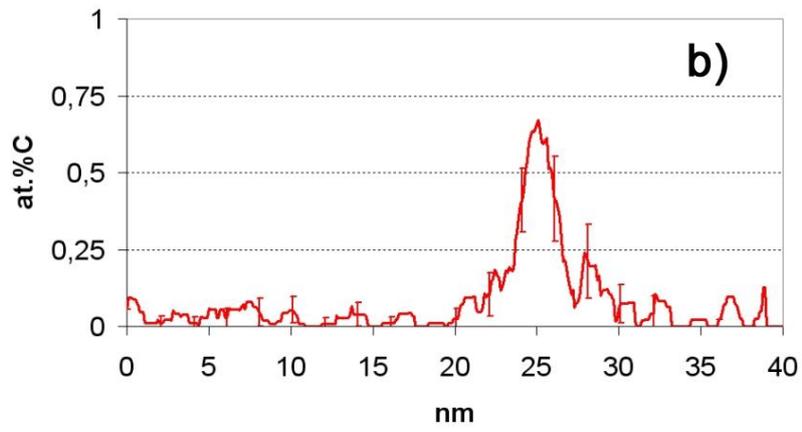

Figure 4